\documentclass[reprint,aps, pre,showpacs, bibnotes, footinbib]{revtex4-1}
\pdfoutput=1
\usepackage{graphicx}
\usepackage{graphicx,amsfonts}
\usepackage{amsmath}
\usepackage{verbatim}

\usepackage{natbib}
\usepackage{textcomp}
\usepackage{amssymb}
\usepackage{float}
\usepackage{footmisc}

\usepackage{color}
\definecolor{darkblue}{rgb}{0,0,0.5}
\definecolor{darkred}{rgb}{0.5,0,0}
\usepackage[colorlinks=true,urlcolor=darkblue,citecolor=darkblue,linkcolor=darkred,hyperfootnotes=false]{hyperref}

\DeclareSymbolFont{bbold}{U}{bbold}{m}{n}
\DeclareSymbolFontAlphabet{\mathbbold}{bbold}

\newcommand{\m}[1]{\mathcal{#1}}
\newcommand{\subte}[2]{#1_{\text{#2}}}

\newcommand{\Erfc}{\text{erfc}}
\begin{document}

\title{Symmetry for the duration of entropy-consuming intervals}

\author{Reinaldo Garc\'ia-Garc\'ia}
\email{reinaldomeister@gmail.com}
\affiliation{Centro At\'omico Bariloche and Instituto Balseiro, 8400 S. C. de Bariloche, Argentina}
\author{Daniel Dom{\'i}nguez}
\affiliation{Centro At\'omico Bariloche and Instituto Balseiro, 8400 S. C. de Bariloche, Argentina}
\begin{abstract}
 We introduce the violation fraction $\upsilon$ as the cumulative fraction of time that a 
 mesoscopic system spends consuming entropy at a single trajectory in phase space. We show that
 the fluctuations of this quantity are described in terms of a symmetry relation reminiscent of fluctuation
 theorems, which involve a function, $\Phi$, which can be interpreted as an entropy associated to the fluctuations
 of the violation fraction.
 The function $\Phi$, when evaluated for arbitrary stochastic realizations of the violation fraction,
 is odd upon the symmetry transformations which are relevant for the associated stochastic entropy production. This fact
 leads to a detailed fluctuation theorem for the probability density function of $\Phi$.
 We study the steady-state limit of this symmetry in the paradigmatic case of a colloidal particle dragged by optical
 tweezers through an aqueous solution. Finally, we briefly discuss on possible applications of our results for the estimation
 of free-energy differences from single molecule experiments.
\end{abstract}
\pacs{05.40.-a,05.70.Ln} 
\maketitle
\section{Introduction}
\label{sec:intro}
Stochastic thermodynamics is, at these days, a very active field given
its great relevance for physics, chemistry, and biology~\cite{Seifert-Review}.
As a consequence of fluctuations, which are extremely important in mesoscopic systems,
the energy change, work, heat, and entropy production associated to any process exhibit
a stochastic nature, i.e., their values are random quantities which depend 
on the particular trajectory the system follows in phase space. 
The fluctuations of some thermodynamic observables are constrained
to satisfy general relations which are known as fluctuation theorems~\cite{Evans-Cohen-Morris,
Gallavotti-Cohen,Kurchan,Lebowitz,Jarzynski-a,*Jarzynski-b,
Crooks-a,*Crooks-b,Hatano-Sasa}, which arise as a consequence of two main properties:
ergodicity and microreversibility. These relations can be generally written as follows
\begin{equation}
 \label{DFT}
\ln\frac{P(S)}{P^{\m{T}}(-S)}=S,
\end{equation}
where $P(S)$ is the probability density function of the stochastic observable $S$
representing a given form of trajectory-dependent entropy production, while
$\m{T}$ represents a transformation, usually time-reversal, the transformation to a {\it dual} dynamics,
or the composition of these two operations
(see~\cite{Jarzynski1,us-unifying,us-joint} for a simple definition of the dual dynamics).
The quantity $S$ exhibits the symmetry $S^{\m{T}}=-S$.

In the last few years, the applicability of these relations have also been extended  
to systems exhibiting stochastic 
non-Markovian dynamics~\cite{Seifert-non-Markov,
Ohkuma-Ohta,Mai-Dhar,BiroliFTs,Zamponi,Klages,me-non-Markov}. On the other hand, they have been widely tested
in experiments~\cite{Wang,Trepagnier,bustamante,carberry,ritort,Gupta}. Fluctuation theorems are also consistent:
the second law of thermodynamics, the Green-Kubo formula, and the Onsager reciprocity relations, for
example, can be deduced from them.

The stochastic nature of thermodynamic observables may lead to a negative value of the entropy production
for particular realizations of a given process. In fact, the occurrence of such rare
realizations is exponentially less probable than the occurrence of trajectories compatible
to the second law of thermodynamics (i.e. realizations where the entropy production is positive), 
as can be immediately interpreted from Eq. (\ref{DFT}). 
It is, however, important to remark that the stochastic entropy production is not restricted
to be positive, i.e. the occurrence of negative values of $S$ at a single trajectory in phase space
does not represent a violation of the second law of thermodynamics, which states that the average
$\langle S\rangle$ is non-negative.

In other matters, if we focus on a 
single trajectory of the system in phase space, one finds that for certain time intervals
the entropy production is negative. 
The statistics of the cumulative duration of these time intervals
has been recently studied~\cite{garcia2012long}. There,  
this duration (relative to the total duration of the process) was denominated as {\it violation fraction},
and the occurrence of negative values of the entropy production as {\it local violations of the second law of thermodynamics},
following the terminology of Ref.~\cite{Wang} (the term 'local' meaning at a single trajectory and stating
that not true violations of the second law occur).
The study made in Ref.~\cite{garcia2012long} is, however, incomplete. First, instead of the full
probability density function of the violation fraction, only its first moment was studied.
Second, some particular conditions were assumed, say, the system was assumed to
be prepared in a steady state and to be connected to a single reservoir. 

The aim of the present paper
is to fill these gaps. 
On these lines, we derive in this work a general symmetry relation for the probability density 
function of the violation fraction.
This relation is valid under very general conditions: the system may be prepared in an arbitrary initial 
state and connected to one or several thermal baths.
No special assumptions are needed  for the results derived in this paper to hold, but only ergodicity and 
microreversibility~\footnote{This statement is valid if we consider Markovian dynamics. When considering non-Markov
dynamics, ergodicity and microreversibility would not be enough. At least \emph{stability} should be 
additionally demanded, as discussed in Ref.~\cite{me-non-Markov}.}.
Our results are rather general and also hold for arbitrary external protocols.

This symmetry relation is reminiscent of fluctuation theorems, and  it involves
a function $\Phi$ which reflects the asymmetry between
forward and transformed processes in phase space as regards the violation fraction. This function satisfies, by construction, 
respective integral and detailed fluctuation theorems, and it is also odd upon some relevant transformations, 
as the entropy production does. Thus, it may be in principle interpreted as an entropy. 
On the other hand, the average of $\Phi$ represents a lower bound for the average
of the entropy production, i.e., the inequality,
$\langle S\rangle\ge\langle\Phi\rangle\ge0$, holds quite generally.

At this point we would like to clarify a subtle but important
issue. In Ref.~\cite{garcia2012long} the violation fraction (denoted by $\nu$ there) was defined
in terms of the entropy produced from the beginning of the process, $S$, i.e. 
the cumulative fraction of time the entropy production from
the beginning of the process was negative. This definition is not invariant upon time
translations, i.e., measuring the entropy production w.r.t. different time instants leads to different 
sets of violation intervals. Here, we consider the stochastic entropy production rate, 
$\sigma=dS/dt$, instead of the stochastic entropy production in order to define the local violations of the second law. 
Then, within the present framework, a violation sector is defined as a time interval
where $\sigma<0$.

The rest of the paper is organized as follows. In the next section we derive a symmetry relation for
the probability density function of the violation fraction, both, in the transient, and in the
stationary regimes. We also discuss the main properties and 
deepen on the physical interpretation of the function $\Phi$ involved in the referred symmetry. 
In section~\ref{sec:case} we study the steady-state limit of the referred symmetry for a
paradigmatic model system. We determine the large-deviation function associated to the violation
fraction in that limit, providing both, particular analytical, and general numerical results.
We finally discuss about the relevance of our study and about possible applications in the estimation
of free energies from single molecule experiments We also provide some concluding remarks
and perspectives of our results, all this in section \ref{sec:conclusion}.

\section{Symmetry for the probability density function of the violation fraction}
\label{sec:symmetry}
\subsection{Transient symmetry}
We start by anticipating the main result of the present section. If we introduce the probability density function for the violation 
fraction $\upsilon(\tau)$ (see Eq. (\ref{viol-def})) to be in the vicinity of the value 
$\upsilon$ at time $\tau$ for the forward ($\rho(\upsilon,\tau)$) 
and transformed ($\rho^{\m{T}}(\upsilon,\tau)$) processes (see details below), we obtain the symmetry relation
\begin{equation}
  \label{main-1}
\ln\frac{\rho(\upsilon,\tau)}{\rho^{\m{T}}(1-\upsilon,\tau)}=\Phi(\upsilon,\tau).
 \end{equation}
The function $\Phi$ is odd upon the corresponding transformation $\m{T}$.
Moreover, making a parallel with classical thermodynamics, $\Phi$ can be seen as an entropy
associated to the local violations of the second law, a sort of ``super-entropy'', as we discuss below. 
Already at this point
we note from Eq.~(\ref{main-1}) that the integral
fluctuation theorem, 
\begin{equation}
 \label{IFT}
\langle e^{-\Phi(\upsilon,\tau)}\rangle=
\int_0^1 d\upsilon\rho(\upsilon,\tau)e^{-\Phi(\upsilon,\tau)}=1, 
\end{equation}
holds at all times $\tau$. From this, and
Jensen's inequality, $-\ln x\ge1-x$ for $x>0$, we see that a ``second-like'' law of thermodynamics
holds for $\Phi$,
$\langle\Phi(\upsilon,\tau)\rangle\ge0$, a result that
can also be seen by identifying the average of $\Phi$ with the 
positively defined Kullback-Leibler distance between
the distributions $\rho$ and $\rho^\m{T}$, $\langle\Phi(\upsilon,\tau)\rangle=\m{D}_{\text{KL}}(\rho||\rho^\m{T})$,
where
\begin{equation}
 \label{KL-distance}
 \m{D}_{\text{KL}}(\rho||\rho^\m{T})=
 \int_0^1 d\upsilon\rho(\upsilon,\tau)\ln\frac{\rho(\upsilon,\tau)}{\rho^\m{T}(1-\upsilon,\tau)}.
\end{equation}

This second law of thermodynamics for $\Phi$ imposes strong restrictions to the stochastic process
$\upsilon(\tau)$. We remark that some minimal requirements are needed for the entropy production associated to a given process
to satisfy the second law of thermodynamics, ergodicity one of them. Its is thus intriguing
that the fraction of time a process spends consuming entropy also exhibits those requirements.

We focus on systems relaxing to well defined
steady states for constant parameters. It is worth to remark the distinction
between two different types of systems:
those where detailed balance holds and the steady state probability density function corresponds
to the Boltzmann-Gibbs distribution, and those relaxing to a nonequilibrium
steady state (NESS) driven by nonconservative forces and/or special boundary conditions.
It is known~\cite{EspositoPRL} that for the latter family of systems the total entropy production ($\subte{S}{tot}$)
splits in an adiabatic contribution ($\subte{S}{a}$) accounting for the
energy dissipated in maintaining a NESS, and a nonadiabatic
contribution ($\subte{S}{na}$) which accounts for the relaxation to the steady state and for the effect of an
external driving. Each of these three forms of entropy production satisfies a fluctuation theorem 
in terms of the corresponding symmetry
operation (time reversal ($\text{R}$) for $\subte{S}{tot}$, dual transformation ($\dagger$) for
$\subte{S}{a}$, and their composition ($\dagger\circ\text{R}$) for $\subte{S}{na}$), leading to three different
faces of the second law of thermodynamics~\cite{Esposito-Master,*Esposito-Fokker}.
Given that our results are valid for each of these forms of entropy production, we generically write
$\upsilon$, $S$, and $\sigma$ without further specification, understanding the notation
$\m{T}$ as the corresponding symmetry transformation associated to each case.

Let us consider an ergodic and microreversible system driven
by a set of external parameters which we denote by $\lambda_t$. The state of the
system in phase space is denoted by the symbol $m$ which may be a discrete state
or a continuous variable, vector, or field. A trajectory in phase space from $t=0$ to
the final time $t=\tau$ is denoted by a bold symbol $\mathbf{m}=\{m(t)\}_{t=0}^\tau$
while the full time-dependence of the protocol is denoted by
$\lambda_\tau=\{\lambda_t\}_{t=0}^\tau$. Additionally, we introduce the time reversed trajectory,
$\mathbf{m}^{\text{R}}=\{m(\tau-t)\}_{t=0}^\tau$, and protocol,
$\lambda_\tau^{\text{R}}=\{\lambda_{\tau-t}\}_{t=0}^\tau$.
As the evolution is stochastic, one can define the probability weights for trajectories,
$\m{P}_\tau[\mathbf{m};\lambda_\tau]$.
A key aspect of ergodic and microreversible systems is that the trajectory
dependent entropy production may be written as the log-ratio of path probability weights
\begin{align}
 \label{stot}
 \subte{S}{tot}[\mathbf{m},\lambda_\tau] &=
\ln\frac{\m{P}_\tau[\mathbf{m};\lambda_\tau]}{\m{P}_\tau^{\text{R}}[\mathbf{m}^{\text{R}};\lambda_\tau^{\text{R}}]},\\
 \label{sna}
 \subte{S}{na}[\mathbf{m},\lambda_\tau] &=
\ln\frac{\m{P}_\tau[\mathbf{m};\lambda_\tau]}{\m{P}_\tau^{\dagger\circ\text{R}}
[\mathbf{m}^{\text{R}};\lambda_\tau^{\text{R}}]},\\
 \label{sa}
\subte{S}{a}[\mathbf{m},\lambda_\tau] &=
\ln\frac{\m{P}_\tau[\mathbf{m};\lambda_\tau]}{\m{P}_\tau^{\dagger}
[\mathbf{m};\lambda_\tau]}, 
\end{align}
leading to the fluctuation theorems for arbitrary protocols, initial conditions and number of reservoirs~\cite{EspositoPRL}.
On the other hand, given an arbitrary observable $\m{O}[\mathbf{m};\lambda_\tau]$, the following equivalences
hold~\cite{us-joint}
\begin{align}
 \label{gen-sym}
 \langle\m{O}[\mathbf{m};\lambda_\tau]\rangle &=\langle\m{O}[\mathbf{m}^{\text{R}};\lambda_\tau]
 e^{-S^{\text{R}}_{\text{tot}}[\mathbf{m},\lambda_\tau^{\text{R}}]}\rangle^\text{R}\nonumber\\
 &=\langle\m{O}[\mathbf{m}^{\text{R}};\lambda_\tau]
 e^{-S^{\dagger\circ\text{R}}_{\text{na}}[\mathbf{m},\lambda_\tau^{\text{R}}]}\rangle^{\dagger\circ\text{R}}\nonumber\\
 &=\langle\m{O}[\mathbf{m};\lambda_\tau]
 e^{-S^{\dagger}_{\text{a}}[\mathbf{m},\lambda_\tau]}\rangle^\dagger.
 \end{align}
where $\langle\ldots\rangle^\m{T}$ denotes
the average with the transformed probability weight $\m{P}_\tau^\m{T}$, for $\m{T}=(\text{R})$, 
$\m{T}=(\dagger\circ\text{R})$, and $\m{T}=(\dagger)$, respectively.
Let us now  formally introduce the violation fraction for a given trajectory as follows
\begin{equation}
 \label{viol-def}
 \upsilon(\tau)\stackrel{\text{def}}{=}\frac{1}{\tau}\int_0^\tau \Theta(-\sigma(t))dt,
\end{equation}
 where $\Theta(\bullet)$ is the Heaviside step function and we have suppressed, for simplicity in the notation,
 the full dependence on the particular trajectory in phase space and the protocol. A key point for our
 derivations is that the violation fraction satisfies the symmetry
 \begin{equation}
  \label{viol-sym}
  \upsilon^\m{T}(\tau)=1-\upsilon(\tau),
 \end{equation}
which can be seen as a direct consequence of microreversibility and ergodicity, similarly as the 
symmetries that the different forms of entropy production exhibit upon the corresponding 
operations~\footnote{For the adiabatic entropy production rate, 
Eq. (\ref{viol-sym}) holds since  $S(\tau)=-S^\dagger(\tau)$
and $S(\tau)=\int_0^\tau dt\sigma(t)$, implying that $\sigma(t)=-\sigma^\dagger(t)$. In the remaining cases,
where a time-reversal operation is involved, Eq. (\ref{viol-sym}) follows from recalling that the entropy production rate is,
for Markov dynamics, a local function of coordinates and protocols 
$\sigma(t)=\sigma(m(t),\dot{m}(t);\lambda(t))$, which changes sign
when evaluated for time reversed arguments.}. 

Let us now introduce the joint probability density function for the violation fraction to exhibit a value in the interval
$[\upsilon,\upsilon+d\upsilon]$ having observed a value of the entropy production in the interval $[S,S+dS]$ at
time $\tau$, $P(\upsilon,S,\tau)$, and the corresponding transformed probability density function,
$P^\m{T}(\upsilon,S,\tau)$, given, respectively, by the following expresions
\begin{align}
 \label{joint-def}
 &P(\upsilon,S,\tau)=\langle\delta(\upsilon-\upsilon(\tau))\delta(S-S(\tau))\rangle,\\
 &P^\m{T}(\upsilon,S,\tau)=\langle\delta(\upsilon-\upsilon^\m{T}(\tau))\delta(S-S^\m{T}(\tau))\rangle^\m{T}.
\end{align}
Using the previous definitions, Eqs~(\ref{gen-sym}) and (\ref{viol-sym}),
and recalling that $S$ is odd upon the operation $\m{T}$, we can write
\begin{equation}
 \label{joint-sym}
 P(\upsilon,S,\tau)=P^\m{T}(1-\upsilon,-S,\tau)e^{S}.
\end{equation}
Our main result, Eq.~(\ref{main-1}), follows by identifying
the probability density function for the violation fraction to be in a vicinity of the value $\upsilon$ at time $\tau$, as
$\rho(\upsilon,\tau)=\int dSP(\upsilon,S,\tau)$, while the transformed probability density function is given by
\begin{equation}
 \label{steps-1}
\rho^{\m{T}}(1-\upsilon,\tau)=\int dSP^\m{T}(1-\upsilon,S,\tau)=\int dSP(\upsilon,S,\tau)e^{-S},
\end{equation}
where we have used a change of variables $S\rightarrow-S$ and the symmetry Eq.~(\ref{joint-sym}).
Introducing now the conditional
probability density function for the entropy production to be in a vicinity of $S$ at time $\tau$ 
given that the observed value of the violation fraction
was $\upsilon$, $P(S,\tau|\upsilon)=P(\upsilon,S,\tau)/\rho(\upsilon,\tau)$, we obtain
Eq.~(\ref{main-1}), with the identification
\begin{equation}
 \label{main-2}
 \Phi(\upsilon,\tau)=-\ln\langle e^{-S(\tau)}|\upsilon\rangle
=-\ln\int dSP(S,\tau|\upsilon)e^{-S}.
\end{equation}

Before closing this subsection, we would like to remark some important issues. First, we note that the existing
fluctuation theorems are exact relations for the probability density function of stochastic observables 
$\m{O}$ exhibiting the symmetry relation $\m{O}^\m{T}=-\m{O}$. The observable considered here, the 
violation fraction, exhibits in contrast a symmetry of the form $\m{O}^\m{T}=1-\m{O}$. If instead of considering
the violation fraction, Eq. (\ref{viol-def}), we study the related 'magnetization'~\cite{dornic1998large} 
$\psi(\tau)$ given by
\begin{equation}
 \label{magnetization}
 \psi(\tau)\stackrel{\text{def}}{=}\frac{1}{\tau}\int_0^\tau \text{Sgn}\big(\sigma(t)\big)dt,
\end{equation}
where $\text{Sgn}(\bullet)$ is the sign function, we recover an observable which is odd upon the
generic transformation $\m{T}$. In this case, we can easily obtain a symmetry relation for $\psi$
using our main result Eq. (\ref{main-1}) and the trivial identity $\psi=1-2\upsilon$.

We note, however, that the referred symmetry
could also be derived from first principles by following the same lines of reasoning leading to (\ref{main-1}).
It is important to remark that 
the violation fraction and the 'magnetization'
are time-averaged quantities, thus,
the symmetry for the probability density function of $\psi$ is a stochastic version of
the functional Crooks theorem derived in Ref.~\cite{williams2011nonequilibrium}
for time-averages of arbitrary phase-space functions in the case of deterministic dynamics starting at equilibrium. 
In our case the observable is stochastic as well as the evolution of the system in phase space, and we consider arbitrary
initial conditions and symmetry transformations.

\subsection{Steady-state symmetry}

When the system asymptotically relaxes to a NESS, 
such that the mean value of the entropy production
rate tends to a constant value, one can derive a steady-state symmetry for the probability
density function of the violation fraction in the same way that a steady-state fluctuation theorem
holds for the entropy production. This is relevant, for example,
when considering small Brownian motors operating under steady-state conditions, or
systems relaxing to a NESS after a fast quench. Other interesting case which falls in
this category is that of a system driven at a constant rate, for which the nonadiabatic
entropy production rate approaches a constant value at large times. The necessary condition
is that the limit
\begin{equation}
 \label{limit-sigma}
 \lim_{\tau\rightarrow\infty}P(\sigma,\tau)=P_\infty(\sigma),
\end{equation}
exists unambiguosly, where, as usual, $\sigma$ could be any particular form of entropy production rate.

Let us introduce the quantity $R(\tau)=\text{Prob}[\sigma(\tau)<0]$. From Eq. (\ref{viol-def}), we have
\begin{equation}
 \label{mean-viol}
 \langle\upsilon(\tau)\rangle=\frac{1}{\tau}\int_0^\tau R(t)dt.
\end{equation}
Assuming that there is a finite characteristic relaxation time $\tau_r$ to reach the NESS, one can see that one has
\begin{align}
 \label{limit-upsilon}
 \langle\upsilon(\tau\gg\tau_r)\rangle &=\frac{1}{\tau}\int_0^{\tau_r} R(t)dt+
 \frac{1}{\tau}\int_{\tau_r}^\tau R(t)dt\nonumber\\
 &\approx\frac{1}{\tau}\int_0^{\tau_r} R(t)dt+R_\infty\bigg(1-\frac{\tau_r}{\tau}\bigg)\nonumber\\
 &=\upsilon_\infty+\frac{A_\infty}{\tau},
\end{align}
to the first nonvanishing order, where $\upsilon_\infty=R_\infty$, $A_\infty=\int_0^{\tau_r} R(t)dt-R_\infty\tau_r$, and
\begin{equation}
 \label{limit-R}
 R_\infty=\lim_{\tau\rightarrow\infty}R(\tau)=\int_{-\infty}^0P_\infty(\sigma)d\sigma.
\end{equation}
We remark here the finite value of $\langle\upsilon(\infty)\rangle=\upsilon_\infty$, in constrast to the 
case studied in Ref.~\cite{garcia2012long} in terms of the entropy production $S$ instead of the entropy
production rate. In that case, it was shown that the average violation fraction vanishes for $\tau\rightarrow\infty$.
However, it is worth noting that to the leading order both quantities relax as $\tau^{-1}$ for large $\tau$, except,
for instance, in the vicinity of a critical point where the large fluctuations may lead to a different asymptotic
behavior~\cite{garcia2012long}.

Additionally, we remark that for $\tau\rightarrow\infty$ the violation fraction converges in
density to its mean, since the system is ergodic:
\begin{align}
 \label{ergo}
 \lim_{\tau\rightarrow\infty}\upsilon(\tau) &=\lim_{\tau\rightarrow\infty}
 \frac{1}{\tau}\int_{0}^\tau \Theta\big(-\sigma(t)\big)dt\nonumber\\
 &=\langle\Theta\big(-\sigma(t)\big)\rangle_\text{ss}\equiv\upsilon_\infty,
\end{align}
where $\langle\ldots\rangle_\text{ss}$ represents the average in the steady state.

To continue, we note that if the system is asymptotically stationary, one expects that the large-time behavior
of the probability density function of the violation fraction may be well described in 
terms of a large deviation function $\zeta(\upsilon)$:
\begin{equation}
 \label{ldf-1}
 \rho(\upsilon,\tau)\sim e^{-\zeta(\upsilon)\tau}.
\end{equation}
The same behavior is expected for $\Phi$, which means that we can write
$\Phi(\upsilon,\tau)\rightarrow\phi(\upsilon)\tau$ for $\tau\rightarrow\infty$, with
\begin{equation}
 \label{ldf-2}
 \phi(\upsilon)=-\lim_{\tau\rightarrow\infty}\frac{1}{\tau}\ln\langle e^{-S(\tau)}|\upsilon\rangle.
\end{equation}
From the previous reasoning, given that in the stationary limit the operation $\m{T}$ is meaningless,
we obtain the steady-state symmetry
\begin{equation}
 \label{ldf-3}
 \zeta(1-\upsilon)-\zeta(\upsilon)=\phi(\upsilon).
\end{equation}

\subsection{Physical properties of the function $\Phi$}

Let us study in more detail the main physical properties of the function $\Phi$ for
arbitrary systems submitted to arbitrary protocols.
The interpretation of this function is by no means exclusive. This means that, if we consider
any other functional of trajectories in phase space instead of the violation fraction, the corresponding
asymmetry function will share the same general properties of $\Phi$.
However, given that the violation fraction measures how likely the comsumption of entropy is,
this characterization is relevant.

We start by noting that, from the definition given by Eq. (\ref{main-2}),
the function $\Phi$ is related to the average of $e^{-S}$ restricted to those
trajectories with fixed value of the violation fraction $\upsilon$. Then, we can establish
a link with classical thermodynamics which clarifies the physical meaning of this function. Note that,
identifying $\upsilon$ with an ``energy'', and $S$ with a ``coordinate'', the conditional probability
$P(S,\tau|\upsilon)$ can be seen as a microcanonical distribution (where energy is fixed), and the
conditional average $\langle e^{-S}|\upsilon\rangle$ as a sort of `` inverse phase space volume'' at fixed
energy. Then, $\Phi(\upsilon,\tau)$ is the microcanonical entropy linked to the energy $\upsilon$.
In this case, the ``equal a priori probability'' postulate needs of the weight $e^{-S}$, which serves as a balance,
since those ``microstates'' with negative values of $S$ (i.e., those trajectories producing negative entropy)
are exponentially less probable.

We now continue by proving that the average of $\Phi$ represents a lower bound for the entropy production.
First, note that the definition given by Eq. (\ref{main-2}) can be rewritten as
\begin{equation}
 \label{strong-1}
 \langle e^{-(S-\Phi(\upsilon,\tau))}|\upsilon\rangle=1.
\end{equation}
From the previous expresion, the following conditional inequality holds:
\begin{equation}
 \label{strong-2}
 \langle S(\tau)|\upsilon\rangle\ge\Phi(\upsilon,\tau)
\end{equation}
Multiplying both terms of Eq. (\ref{strong-2}) by $\rho(\upsilon,\tau)$ and integrating $\upsilon$ out, we obtain
\begin{equation}
 \label{strong-3}
 \langle S(\tau)\rangle\ge\langle\Phi(\upsilon,\tau)\rangle.
\end{equation}

It is straightforward to see that $\Phi$ is odd upon the corresponding operation $\m{T}$.
Indeed, introducing the transformed potential
$\Phi^\m{T}(\upsilon,\tau)=-\ln\langle e^{-S}|\upsilon\rangle^\m{T}$, we have
\begin{align}
 \label{sym-phi}
 \Phi^\m{T}(1-\upsilon,\tau) &=-\ln\int dSP^\m{T}(S,\tau|1-\upsilon)e^{-S}\nonumber\\
 &=-\ln\int dS\frac{P^\m{T}(1-\upsilon,S,\tau)}{\rho^\m{T}(1-\upsilon,\tau)}e^{-S}\nonumber\\
 &=-\ln\int dS\frac{P^\m{T}(1-\upsilon,-S,\tau)}{\rho(\upsilon,\tau)}e^{S}e^{\Phi(\upsilon,\tau)}\nonumber\\
 &=-\Phi(\upsilon,\tau)-\ln\int dSP(S,\tau|\upsilon)\nonumber\\
 &\equiv-\Phi(\upsilon,\tau),
\end{align}
where we have used Eqs.~(\ref{main-1}), (\ref{joint-sym}), a change of variables
$S\rightarrow-S$, the definition of the conditional
probability and its normalization condition $\int dSP(S,\tau|\upsilon)=1$. Let us introduce
the probability density function for the values of $\Phi$, $P(\Phi,\tau)$, and its transformed counterpart, $P^\m{T}(\Phi,\tau)$,
as follows:
\begin{align}
\label{P-phi}
P(\Phi,\tau) &=\int_0^1d\upsilon\delta(\Phi-\Phi(\upsilon,\tau))\rho(\upsilon,\tau),\\
\label{P-phi-T}
P^\m{T}(\Phi,\tau) &=\int_0^1d\upsilon\delta(\Phi-\Phi^\m{T}(\upsilon,\tau))\rho^{\m{T}}(\upsilon,\tau).
\end{align}
Then, using the previous definitions complemented by Eqs. (\ref{main-1}) and (\ref{sym-phi}), we obtain that
a detailed fluctuation theorem also holds for $\Phi$:
\begin{equation}
 \label{DFT_Phi}
 \ln\frac{P(\Phi,\tau)}{P^\m{T}(-\Phi,\tau)}=\Phi.
\end{equation}
All these properties strongly support the interpretation of $\Phi$ as an entropy associated to the
fluctuations of the violation fraction.

Having defined the function $\Phi$ by Eq. (\ref{main-2}), we now derive an alternative formula to determine
this quantity which provides more insight on its physical meaning. If we divide both terms of 
Eq. (\ref{joint-sym}) by $\rho(\upsilon,\tau)$ and use Eq. (\ref{main-1}), we obtain
\begin{equation}
 \label{alternative-phi-1}
 P(S,\tau|\upsilon)=P^\m{T}(-S,\tau|1-\upsilon)e^{S-\Phi(\upsilon,\tau)},
\end{equation}
from where we immediately get
\begin{equation}
 \label{alternative-phi-2}
 \Phi(\upsilon,\tau)=S-\ln\frac{P(S,\tau|\upsilon)}{P^\m{T}(-S,\tau|1-\upsilon)}.
\end{equation}
The previous expresion is valid for any value of $S$, but it turns into a very meaningful formula
when we consider the case of $S=0$:
\begin{equation}
 \label{alternative-phi-3}
 \Phi(\upsilon,\tau)=-\ln\frac{P(0,\tau|\upsilon)}{P^\m{T}(0,\tau|1-\upsilon)}.
\end{equation}

This alternative definition with a focus on trajectories which do not produce entropy,
provides a clear interpretation of $\Phi$ and allows us to show that $\Phi(1/2,\tau)\equiv0$ irrespectively
of the particular protocols and of the value of $\tau$, as we discuss below.

Let us denote the
full space of possible trajectories in phase space 
as $\Omega$, and let us introduce
the set $\mathbb{V}_\lambda=\{\mathbf{m}\in\Omega|\upsilon(\tau)=1/2\}$, for 
a given protocol $\lambda_\tau$, which can be arbitrarily chosen. Let us also
introduce the 'twin' set $\mathbb{V}_{\lambda^\m{T}}$, corresponding to the transformed
protocol $\lambda_\tau^\m{T}$ and the transformed dynamics.  
It is worth noting that, by virtue of Eq. (\ref{viol-sym}), the set $\mathbb{V}_\lambda$
maps to $\mathbb{V}_{\lambda^\m{T}}$ under the transformation $\m{T}$. Indeed, given that $\upsilon=1/2$ if
and only if $\upsilon=\upsilon^\m{T}$, we see that for any trajectory $\mathbf{m}\in\mathbb{V}_\lambda$,
we have that $\mathbf{m}^\m{T}\in\mathbb{V}_{\lambda^\m{T}}$ also. On the other hand,
if $\mathbf{m}\notin\mathbb{V}_\lambda$, then $\mathbf{m}^\m{T}\notin\mathbb{V}_{\lambda^\m{T}}$
either. 

This, however, does not hold for arbitrary subsets of $\mathbb{V}_\lambda$. In particular, let us
introduce the parametrized family of subsets 
$\mathbb{S}_\lambda(S)$ of $\mathbb{V}_\lambda$ as
$\mathbb{S}_\lambda(S)=\{\mathbf{m}\in\mathbb{V}_\lambda|S(\tau)=S\}$, and the corresponding family
under the transformed dynamics,  
$\mathbb{S}_{\lambda^\m{T}}(S)=\{\mathbf{m}\in\mathbb{V}_{\lambda^\m{T}}|S(\tau)=S\}$.
Then, given that
the entropy production satisfies the symmetry $S^{\m{T}}=-S$, we have that for any $S$,
$\mathbb{S}_\lambda(S)$ maps to $\mathbb{S}_{\lambda^\m{T}}(-S)$ under $\m{T}$:
\begin{equation}
 \label{map-S}
 \mathbb{S}_\lambda(S)\stackrel{\m{T}}{\longmapsto}\mathbb{S}_{\lambda^\m{T}}(-S).
\end{equation}
A special case is that one for which $S=0$, because, from Eq. (\ref{map-S}) we see 
that $\mathbb{S}_\lambda(0)\stackrel{\m{T}}{\longmapsto}\mathbb{S}_{\lambda^\m{T}}(0)$. In particular, if
$\mathbf{m}\in\mathbb{S}_\lambda(0)$, and its probability weight is $\m{P}[\mathbf{m};\lambda]$, it
is easy to see that $\mathbf{m}^\m{T}\in\mathbb{S}_{\lambda^\m{T}}(0)$ also, and that the probability weight
of $\mathbf{m}^\m{T}$ is also $\m{P}[\mathbf{m};\lambda]$ 
(i.e., $\m{P}[\mathbf{m};\lambda]=\m{P}^\m{T}[\mathbf{m}^\m{T};\lambda^\m{T}]$) because $S=0$. Then, we may write:
\begin{equation}
 \label{important-ratio}
 P(0,\tau|1/2)=P^\m{T}(0,\tau|1/2).
\end{equation}
Evaluating Eq. (\ref{alternative-phi-3}) for $\upsilon=1/2$ and using Eq. (\ref{important-ratio}),
we immediately obtain $\Phi(1/2,\tau)=0$.

We finish this section by noting that,
by virtue of the second law of thermodynamics, the small values of 
the violation fraction are more likely than the large values of this quantity.
For small values of $\upsilon$, $\Phi$ is positive, while for large values of $\upsilon$,
$\Phi$ is negative. We conjecture that $\Phi(\upsilon,\tau)$ must quite generally be a decreasing function of $\upsilon$
for $\upsilon\in[0,1]$, which means that $\upsilon=1/2$ is the only zero of $\Phi$, and that
$\Phi(\upsilon,\tau)$ admits an inverse function.

\section{Large deviation function for the violation fraction: A case study}
\label{sec:case}

\subsection{Posing of the problem}

In this section we study and determine the large-deviation function
of the violation fraction in the paradigmatic case of an overdamped
colloidal particle dragged through a viscous fluid by an
optical tweezer with harmonic potential
\begin{equation}
 \label{tweezer}
 V(x;\lambda)=\frac{1}{2}(x-\lambda)^2,
\end{equation}
where the focus of the optical tweezer is moved at a constant rate $b$, $\lambda(t)=bt$.
Although being widely studied, this example is still instructive. On the other
hand, even in this simple case the derivation of a closed analytical solution is out of scope.
The system evolves under the Langevin dynamics
\begin{equation}
 \label{Lang-evol}
 \dot{x}(t)=-(x(t)-bt)+\sqrt{2T}\xi(t),
\end{equation}
where the white noise $\xi(t)$ has zero mean and variance $\langle\xi(t)\xi(t')\rangle=\delta(t-t')$.
If the system is initially prepared in the steady state associated to $\lambda(0)=0$, the stochastic
entropy production corresponds in this case to the Jarzynski work~\cite{Jarzynski-a,*Jarzynski-b}\footnote{An additional
term involving the change in free-energy, $\Delta F(t)$, is generally present, however, in this case
the free energy does not depend on $\lambda$ and one only needs to consider the thermodynamic work.}:
\begin{equation}
 \label{work}
 S(t)=\int_0^t\dot{\lambda}(t')\partial_{\lambda}V(x(t');\lambda(t'))dt',
\end{equation}
from where the stochastic entropy production rate can be identified as
\begin{equation}
 \label{EPR}
 \sigma(t)=\frac{b}{T}(bt-x(t)).
\end{equation}
Let us introduce a new stochastic process $\eta(t)$ as 
\begin{equation}
 \label{effective}
 \eta(t)=\frac{1}{\sqrt{2T}}(bt-x(t))-\eta_m,
\end{equation}
with $\eta_m=b/\sqrt{2T}$.
Then, the equation of motion for this process reads
\begin{equation}
 \label{eff-motion}
 \dot{\eta}(t)=-\eta(t)+\xi(t).
\end{equation}
Note that one has that $\sigma(t)<0$ if and only if $\eta+\eta_m<0$, which means
that the violation fraction for our problem can be written in terms of the auxiliary
process $\eta(t)$ as
\begin{equation}
 \label{viol-eff}
 \upsilon(\tau)=\frac{1}{\tau}\int_0^\tau\Theta(-\eta(t)-\eta_m)dt.
\end{equation}
The statistics of the occupation times associated to the 
Ornstein-Uhlenbeck process given by Eq. (\ref{eff-motion})
have been widely studied in the literature (see for instance Ref.~\cite{densing2012occupation}).
Furthermore, there is a well-established method to compute the large deviation function
associated to any non-linear functional of $\eta$~\cite{majumdar2002large,bray2013persistence}.
We briefly review the method below, as presented in Ref.~\cite{majumdar2002large}. 

\subsection{Large-deviation function for arbitrary time-averaged quantities}

The probability distribution of the process $\eta(t)$ for
$0\le t\le\tau$ is given by
\begin{equation}
 \label{OU-prob}
 \m{P}[\eta]=\m{N}\exp\bigg\{-\frac{1}{2}\int_0^\tau\big[\dot{\eta}(t)+\eta(t)\big]^2dt\bigg\},
\end{equation}
where $\m{N}$ is a normalization constant. We are interested on the probability density function
of the time-averaged quantity
\begin{equation}
 \label{r-def}
 r(\tau)=\frac{1}{\tau}\int_0^\tau U_0\big(\eta(t)\big)dt,
\end{equation}
where $U_0(\eta)$ is an arbitrary function of the stochastic variable $\eta$.
In practice it is convenient to look at the
distribution $P_u(u)$ of the quantity $u=r\tau$. Its Laplace transform reads
\begin{equation}
 \label{Laplace}
 \hat{P}_u(s)=\langle\exp(-rs\tau)\rangle=Z(s)/Z(0),
\end{equation}
with
\begin{equation}
 \label{Z-def}
 Z(s)=\int\m{D}[\eta]\exp\bigg\{-\frac{1}{2}\int_0^\tau\big[\dot{\eta}^2+2\eta\dot{\eta}+\eta^2+2sU_0(\eta)\big]dt\bigg\}.
\end{equation}
We are interested in the limit $\tau\rightarrow\infty$. It is convenient to
impose periodic boundary conditions, $\eta(\tau)=\eta(0)$, since
this restriction will not change the results in the large-$\tau$ limit.
With this, we can drop the term $2\eta\dot{\eta}$, which is a perfect derivative.
Then, $Z(s)$ is the
imaginary-time Feynman path integral that gives the partition function of a quantum particle with Hamiltonian 
$H=p^2/2+\eta^2/2+sU_0(\eta)$ at inverse temperature $\tau$, $p$ being the
canonical momentum conjugate to $X$. For $\tau\rightarrow\infty$ the ground
state dominates:
\begin{equation}
 \label{ground}
 \langle\exp(-rs\tau)\rangle=\exp\big\{-\tau\big[E_g(s)-E_g(0)\big]\big\},
\end{equation}
where $E_g(s)$ is the ground-state energy for the Schr\"odinger
equation
\begin{equation}
 \label{schrodinger}
 -\frac{1}{2}\frac{d^2\psi(\eta)}{d\eta^2}+U(\eta,s)\psi(\eta)=E(s)\psi(\eta),
\end{equation}
with 
\begin{equation}
 \label{new-U}
U(\eta,s)=\frac{\eta^2}{2}+sU_0(\eta).
\end{equation}
For $s=0$ the problem reduces to a simple harmonic oscillator, and $E_g(0)=1/2$.
We now note that, from Eq. (\ref{ground}), the large-time behavior of
$P(r,\tau)$ is given by the inverse Laplace transform:
\begin{equation}
 \label{inverse}
 P(r,\tau)\propto\int_{-i\infty}^{i\infty}ds\exp\big[\tau g(s)\big],
\end{equation}
where $g(s)=rs+E_g(0)-E_g(s)$. Using the steepest-descent method, one sees that we
have $P(r,\tau)\approx\exp[-\zeta(r)\tau]$, with
\begin{equation}
 \label{max-condition}
 \zeta(r)=\max_{s}\big[E_g(s)-E_g(0)-rs\big].
\end{equation}
We now use this method to compute the large-deviation function of the violation
fraction associated to the process given by Eq. (\ref{Lang-evol}).

\subsection{Large-deviation function for the violation fraction}

\begin{figure}
\centering
  \includegraphics[scale=0.9]{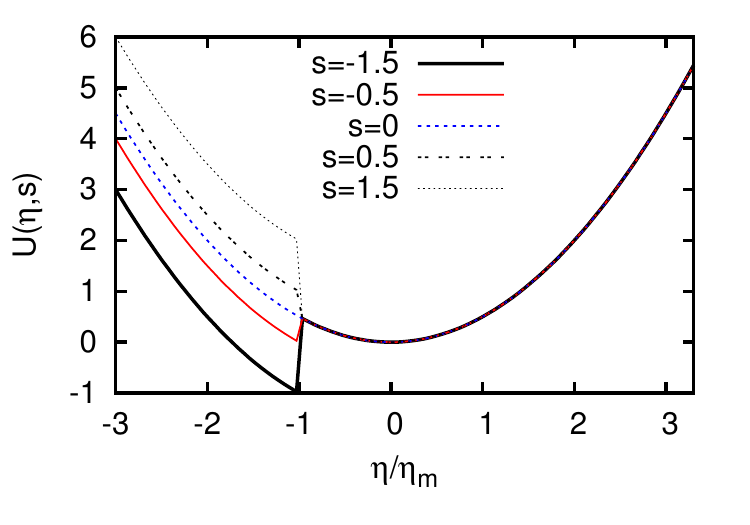}
  \caption{(Color online) Effective potential for equivalent quantum problem associated to the large-deviation
  function of the violation fraction, for different values of $s$.
At $\eta=-\eta_m$, the parabolic potential exhibits a jump of magnitude $U(-\eta_m^+,s)-U(-\eta_m^-,s)=-s$.}
  \label{fig1}
\end{figure}

From Eq. (\ref{viol-eff}), we see that for the violation fraction the effective potential reads
\begin{equation}
 \label{eff-pot-use}
 U(\eta,s)=\frac{\eta^2}{2}+s\Theta(-\eta-\eta_m).
\end{equation}
In Fig.~\ref{fig1} we plot this potential for different values of $s$, to explicitly show its shape.
At $\eta=-\eta_m$, the parabolic potential exhibits a jump of magnitude $-s$.

We now continue by noting that in our problem the Schr\"odinger equation acquires the following particular form:
\begin{align}
 \label{quantum-right}
\psi_+''-\eta^2\psi_++2E(s)\psi_+ &=0,\\
 \label{quantum-left}
\psi_-''-\eta^2\psi_--2s\psi_-+2E(s)\psi_- &=0,
\end{align}
where $\psi_+(\eta)=\psi(\eta>-\eta_m)$, $\psi_-(\eta)=\psi(\eta<-\eta_m)$,
and $\psi''=d^2\psi/d\eta^2$.

As in Ref.~\cite{majumdar2002large}, the solutions of Eqs. (\ref{quantum-right}) and (\ref{quantum-left})
can be expressed in terms of parabolic cylinder functions, $D_p(z)$, using the standard solutions of the
parabolic cylinder equation, $y''-(z^2/4+a)y=0$. Selecting the solutions that satisfy
the physical boundary condition $\psi(\pm\infty)=0$ gives
\begin{align}
 \label{quantum-right-sol}
 \psi_+(\eta) &=AD_{p^+}(\sqrt{2}\eta),\\
 \label{quantum-left-sol}
 \psi_-(\eta) &=BD_{p^-}(-\sqrt{2}\eta),
\end{align}
where $A$ and $B$ are normalization constants while $p^+$ and $p^-$ are given by
\begin{align}
 \label{p-plus}
 p^+ &=E(s)-\frac{1}{2},\\
 \label{p-minus}
 p^- &=E(s)-\frac{1}{2}-s.
\end{align}
Imposing the continuity of $\psi$ and $\psi'$ at $\eta=-\eta_m$, we obtain the following eigenvalue equation
for $E(s)$:
\begin{equation}
 \label{eigenvalue}
 \frac{D_{p^+}'(-\sqrt{2}\eta_m)}{D_{p^+}(-\sqrt{2}\eta_m)}=-
 \frac{D_{p^-}'(\sqrt{2}\eta_m)}{D_{p^-}(\sqrt{2}\eta_m)},
\end{equation}
from where the ground-state energy, $E_g(s)$, and all the excited states can be obtained. 
In general, this problem can not be solved
analytically for generic values of $s$ and $\eta_m$, so, a numerical solution of the eigenvalue
problem, Eq. (\ref{eigenvalue}), is mandatory. Nevertheless, certain interesting limits
can be studied analytically, as we discuss below.

\subsubsection{The limit $\upsilon\rightarrow0$}

In the limit $\upsilon\rightarrow0$ the violation sector,
$\eta<-\eta_m$, is almost inaccessible which means,
appealing to the reader's physical intuition, that this 
scenario is compatible with an infinite wall at $\eta=-\eta_m$.
The present limit corresponds then to the case $s\rightarrow+\infty$.

It has been recently shown that in the case of an infinite wall displaced to the left from the center of a parabolic
potential, the ground-state energy, $\epsilon_0(\eta_m)$, decreases monotonically as a function of 
the center-wall distance~\cite{fernandez2010simple}. Thus, for a hard wall at $\eta=-\eta_m$,
we have $E_g(+\infty)=\epsilon_0(\eta_m)$, with $1/2<\epsilon_0(\eta_m)\le3/2$, and
$d\epsilon_0(\eta_m)/d\eta_m<0$. The ground-state energy, $\epsilon_0(\eta_m)$, has to be 
determined numerically for arbitrary $\eta_m$~\cite{fernandez2010simple,mei1983harmonic}. 

We need the first
correction to this result for large $s$. Although it can be derived formally, we will continue
by using physical arguments to render calculations easier and to enlighten the discussion.

First, note that for $s\rightarrow\infty$ the details of the quadratic potential are not important
in Eq. (\ref{quantum-left}), and we can see the problem as an equivalent problem with a high potential
barrier (of magnitude $s$) for $\eta<-\eta_m$. In this case 
the ground-state wave-function penetrates the barrier
by a typical depth $\delta^+_g(s)=[2(s-E_g(s))]^{-1/2}\approx(2s)^{-1/2}$, where it drops to zero.
The problem is then equivalent, in physical terms, to a problem with an infinite hard wall but
placed at $-\eta_\delta=-\eta_m-\delta^+_g(s)$. Note that in the equivalent model the wave-function
identically vanishes at $\eta=-\eta_\delta$. We can then write to the leading order in $\delta^+_g(s)$:
\begin{equation}
 \label{ground-v0}
 E_g(s)=\epsilon_0(\eta_m)-|\epsilon_0'(\eta_m)|(2s)^{-1/2}.
\end{equation}
The previous equation coincides (without droping the term $E_g(s)\approx E_g(\infty)$ from the square root)
with the formal result obtained in Ref.~\cite{mei1983harmonic} by means of perturbation theory. The
formal equivalence can be easily seen by using the virial theorem given 
by Eq. (15) of Ref.~\cite{fernandez2010simple}. Using now Eq. (\ref{ground-v0}) and the general method
given by Eq. (\ref{max-condition}), we obtain in the limit $\upsilon\rightarrow0$:
\begin{equation}
 \label{ldf-v0}
 \zeta(\upsilon)=\zeta_0-a_0\upsilon^{1/3},
\end{equation}
with $\zeta_0=\epsilon_0(\eta_m)-1/2$, and $a_0=\frac{3}{2}|\epsilon_0'(\eta_m)|^{2/3}$. In the limit of
slow driving, $\eta_m\ll1$, we can write a closed expresion for $\zeta(\upsilon)$ to
the first order in $\eta_m$. Using $\epsilon_0(0)=3/2$, 
and $|\epsilon_0'(0)|=2/\sqrt{\pi}$~\cite{mei1983harmonic}, we have
\begin{equation}
 \label{ldf-v0-slow}
 \zeta(\upsilon)=-\frac{2}{\sqrt{\pi}}\eta_m+\zeta_\text{qs}(\upsilon),
\end{equation}
where the quasistatic large-deviation function, $\zeta_\text{qs}(\upsilon)$, is given by
\begin{equation}
 \label{ldf-v0-qs}
 \zeta_\text{qs}(\upsilon)=1-\frac{3}{2}\bigg(\frac{4}{\pi}\upsilon\bigg)^{1/3},
\end{equation}
which is the expected result for $\eta_m=0$. The last statement can be seen by considering the 'magnetization'
$\psi$. Substituting in Eq. (\ref{ldf-v0-qs}) the identity $\upsilon=(1-\psi)/2$, and noting that
for $\upsilon\rightarrow0$, we have that $\psi\rightarrow1$, we obtain exactly Eq. (60) of Ref.~\cite{majumdar2002large}.

\subsubsection{The limit $\upsilon\rightarrow1$}

We now study the limit of large values of the violation fraction. In this limit, trajectories spend
most of the time in the violation sector, $\eta<-\eta_m$. Then, the right branch of the parabolic potential
is almost inaccessible, a situation which is compatible with the limit $s\rightarrow-\infty$.
In this case, the potential has a deep minimum at $\eta=-\eta_m$, with energy $\eta_m^2/2+s$ thus, it is convenient
to redefine $E(s)=s+\varepsilon(s)$. With this, Eqs. (\ref{quantum-right})-(\ref{quantum-left})
read:
\begin{align}
 \label{quantum-right-shift}
\psi_+''-\eta^2\psi_++2s\psi+2\varepsilon(s)\psi_+ &=0,\\
 \label{quantum-left-shift}
\psi_-''-\eta^2\psi_-+2\varepsilon(s)\psi_- &=0.
\end{align}
We now note that, for $s\rightarrow-\infty$, the details of the quadratic potential are not important
in Eq. (\ref{quantum-right-shift}). Then, just as we did in the limit $\upsilon\rightarrow0$, we can neglect
the effect of the parabolic potential in (\ref{quantum-right-shift}) and consider a high potential
barrier of height $-s$. With this, the ground-state wave-function penetrates the region $\eta>-\eta_m$ by a
small depth $\delta^-_g(s)\approx(-2s)^{-1/2}$, and we can approximate our problem by an equivalent
one with a hard wall at $\eta=-\eta_m+\delta^-_g(s)$. For $\eta<-\eta_m$, the solution of 
Eq. (\ref{quantum-left-shift}) is still given by a parabolic cylinder function
\begin{equation}
 \label{quantum-left-shift-sol}
 \psi_-(\eta)=AD_p(-\sqrt{2}\eta),
\end{equation}
where $A$ is a normalization constant, and $p=\varepsilon(s)-1/2$. The hard wall condition gives the
eigenvalue equation for $\varepsilon(s)$:
\begin{equation}
 \label{left-shift-eigen-1}
 D_p\big(\sqrt{2}(\eta_m-\delta^-_g(s))\big)=0.
\end{equation}
Eq. (\ref{left-shift-eigen-1}) has still to be solved numerically, however, the limit of slow driving,
$\eta_m\ll1$, can be treated analytically. For any small $\eta$, we can write
$D_p(-\sqrt{2}\eta)\approx D_p(0)-\sqrt{2}D'_p(0)\eta$. With this, we can rewrite our eigenvalue
equation as
\begin{equation}
 \label{left-shift-eigen-2}
 \frac{D_p(0)}{D'_p(0)}=\sqrt{2}\eta_m-(-s)^{-1/2}.
\end{equation}
Given that the right hand side of Eq. (\ref{left-shift-eigen-2}) is a small quantity, we can expand
$\varepsilon(s)=3/2-\epsilon$ for the ground-state, with $\epsilon\ll1$. Using standard identities relating the parabolic
cylinder functions to $\Gamma$ functions~\cite{abramowitz2012handbook}, we have
\begin{equation}
 \label{left-shift-eigen-3}
 \frac{\Gamma\big(-\frac{p}{2}\big)}{\Gamma\big(\frac{1-p}{2}\big)}=\sqrt{2}(-s)^{-1/2}-2\eta_m.
\end{equation}
Note that $p=1-\epsilon$. Expanding the $\Gamma$ functions above for small $\epsilon$ we have
$\Gamma\big((\epsilon-1)/2\big)/\Gamma\big(\epsilon/2\big)\approx-\sqrt{\pi}\epsilon$, thus,
we obtain
\begin{equation}
 \label{left-shift-eigen-3}
 \epsilon=\frac{2}{\sqrt{\pi}}\eta_m-\sqrt{-\frac{2}{\pi s}}.
\end{equation}
We then have for the ground-state energy
\begin{equation}
 \label{left-shift-eigen-4}
 E_g(s)=s+\frac{3}{2}-\frac{2}{\sqrt{\pi}}\eta_m+\sqrt{-\frac{2}{\pi s}},
\end{equation}
from where we obtain for the large-deviation function, in the limit $\upsilon\rightarrow1$:
\begin{equation}
 \label{ldf-v1-slow}
 \zeta(\upsilon)=-\frac{2}{\sqrt{\pi}}\eta_m+\zeta_\text{qs}(1-\upsilon),
\end{equation}
with $\zeta_\text{qs}$ again given by Eq. (\ref{ldf-v0-qs}). We point out that, again, for
$\eta_m=0$, and using $1-\upsilon=(1+\psi)/2$ in terms of the magnetization $\psi$, we
obtain Eq. (59) of Ref.~\cite{majumdar2002large}.

We would like to remark that the behavior of $\zeta(\upsilon)$ is, from Eqs. (\ref{ldf-v0-slow})
and (\ref{ldf-v1-slow}), similar for $\upsilon\rightarrow0$ and $\upsilon\rightarrow1$.
On the other hand, one expects these behaviors to be different for nonzero $\eta_m$, since the
second law of thermodynamics favors small values of the violation fraction and penalizes large
values of this quantity. Our analytical results show that the difference in the behaviour around 
$\upsilon\rightarrow0$ with respect to the behavior around $\upsilon\rightarrow1$ is, at least,
of second order of perturbation theory in $\eta_m$, around $\eta_m=0$. Thus for small,
still finite values of $\eta_m$, one expects $\zeta(\upsilon)$ to be very symmetric around $\upsilon=1/2$,
exactly as for $\eta_m=0$.

\subsubsection{General results}

\begin{figure}
\centering
  \includegraphics[scale=0.9]{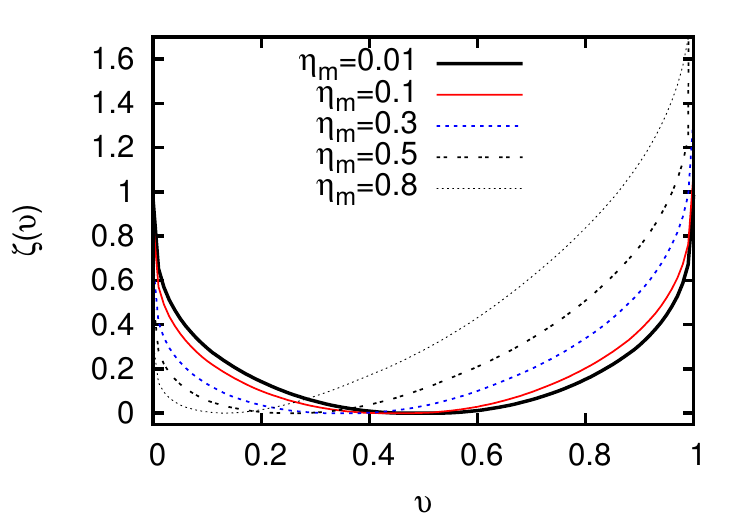}
  \caption{(Color online) Large-deviation function for the violation fraction for 
  $\eta_m=0.01$, $0.1$, $0.3$, $0.5$, and $0.8$. For $\eta_m=0.01$ the large-deviation function
  is very symmetric with respect to $\upsilon=1/2$. The position of the minimum of this function decreases very
  rapidly as $\eta_m$ increases.}
  \label{fig2}
\end{figure}

We now turn to the numerical solution of Eq. (\ref{eigenvalue}). In Fig. \ref{fig2} we plot
the large-deviation function obtained numerically. It can be seen that for $\eta_m=0.01$ the
large-deviation function is very symmetric, a fact that is in concordance with our analytical
results, Eqs. (\ref{ldf-v0-slow}) and (\ref{ldf-v1-slow}). As $\eta_m$ increases, the position of the minimum
of the large-deviation function decreases very rapidly, while this function becomes very asymmetric, even
for $\eta_m<1$. This fact can be understood as follows. From the dynamics given by Eq. (\ref{Lang-evol}),
and the definition of the entropy production rate, Eq. (\ref{EPR}), we obtain that in the stationary limit
the probability density function of $\sigma$ reads
\begin{equation}
 \label{steady-epr}
 P_\infty(\sigma)=\frac{1}{2\sqrt{\pi}\eta_m}\exp\bigg[-\frac{1}{4\eta_m^2}(\sigma-2\eta_m^2)^2\bigg].
\end{equation}

\begin{figure}[H]
\centering
  \includegraphics[scale=0.9]{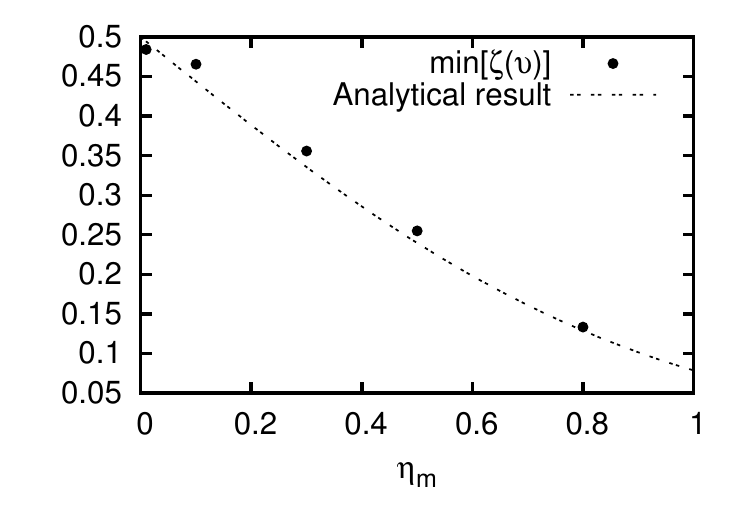}
  \caption{Comparison between the numerically obtained position of the minimum of $\zeta(\upsilon)$
  for different values of $\eta_m$, and the exact result given by Eq. (\ref{steady-viol-mean}). Both
  results coincide within the numerical errors.}
  \label{fig3}
\end{figure}

\begin{figure}
\centering
  \includegraphics[scale=0.9]{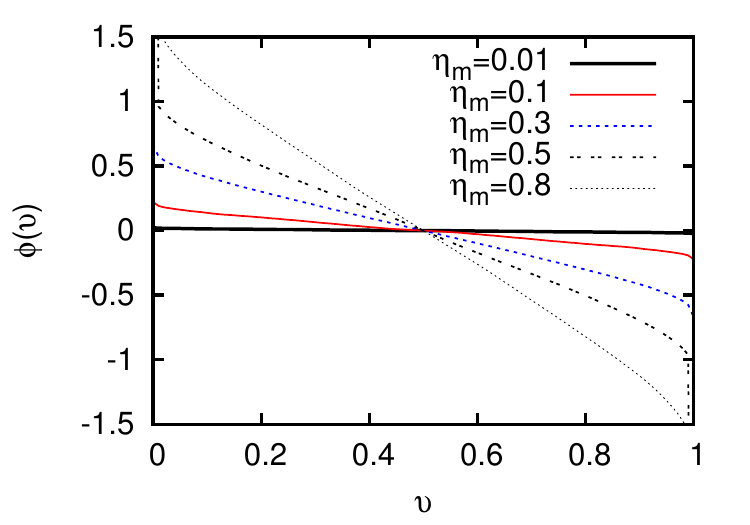}
  \caption{(Color online) Asymmetry function, $\phi(\upsilon)$, for 
  $\eta_m=0.01$, $0.1$, $0.3$, $0.5$, and $0.8$. As $\eta_m$ approaches the value $\eta_m=1$,
  the slope of the asymmetry function rapidly becomes very steep.}
  \label{fig4}
\end{figure}

From this result, we get that the mean value of the violation fraction in this limit, is given by
\begin{equation}
 \label{steady-viol-mean}
 \langle\upsilon(\tau)\rangle=\int_{-\infty}^0P_\infty(\sigma)d\sigma=\frac{1}{2}\Erfc(\eta_m),
\end{equation}

where $\Erfc(\bullet)$ is the complemetary error function. Recalling that the position of the
minimum of the large-deviation function corresponds to $\langle\upsilon\rangle$, this explains
why this point shifts so rapidly to the left when we increase $\eta_m$. In Fig. \ref{fig3} we
plot the numerically obtained position of the minimum of the large-deviation function and the
exact result given by Eq. (\ref{steady-viol-mean}), obtaining a very good agreement between
both results within the numerical errors.
The asymmetry function, $\phi(\upsilon)$, can be obtained directly from Eq. (\ref{ldf-3}). We plot
this function in Fig. \ref{fig4}. For $\eta_m=0.01$ this function is almost flat, since in this case
the thermal fluctuations are large and/or the driving velocity is small (recall the definition of $\eta_m$),
which means that the local violations of the second law are more probable in this limit. As the driving velocity
increases (and/or the temperature decreases), the asymmetry between small and large values of the violation
fraction increases very rapidly. 

This is easy to understand. Note that, from the exact result given by
Eq. (\ref{steady-epr}), we see that $\langle\sigma\rangle=2\eta_m^2$, while 
$\sqrt{2\langle\delta\sigma^2\rangle}=2\eta_m$, with $\delta\sigma(t)=\sigma(t)-\langle\sigma(t)\rangle$.
Then, as long as the amplitude of the fluctuations is greater than $\langle\sigma\rangle$, the local
violations are  likely to occur, but if $\sqrt{2\langle\delta\sigma^2\rangle}<\langle\sigma\rangle$, i.e.,
if $\eta_m>1$, the local violations are extremely rare and the asymmetry function develops a very
steep slope. This physical analysis explains why 
solving Eq. (\ref{eigenvalue}) numerically for $\eta_m\ge1$ and arbitrary values of $s$ is so difficult.
In that region, the numerical scheme implemented by us becomes unstable.

\section{Discussion and concluding remarks}
\label{sec:conclusion}
We have studied the statistics of the ocurrence of entropy-consuming time intervals
for single realizations of stochastic processes in phase space. The conditions for
the validity of the main results derived here are very general, however, we would like
to briefly discuss an important point regarding the definition of the violation fraction
itself, which is the central object in our theory.
The average entropy production rate is quite generally well defined, however, its stochastic counterpart
may be sometimes ill-defined. Note that in certain cases, just in the same way as the velocity of
a Brownian walker, the stochastic entropy production rate is well defined only under an integration sign, i.e, in the
sense of generalized functions.
The stochastic entropy production $S$ is, on the other hand, allways well defined. However, 
one can introduce a 'coarse-grained'
entropy production rate by considering an arbitrarily small, but still finite, time window $\tau_w$.
Generalizing the definition given by Eq. (\ref{viol-def}) to this case, we may write:
\begin{equation}
 \label{viol-coarse}
 \upsilon(\tau,\tau_w)=\frac{1}{\tau-\tau_w}\int_0^{\tau-\tau_w}\Theta\bigg(\frac{S(t)-S(t+\tau_w)}{\tau_w}\bigg)dt.
\end{equation}
The limit $\tau_w\rightarrow0^+$, if it exists, corresponds to the definition given by Eq. (\ref{viol-def}).
It turns out that, using $\upsilon(\tau,\tau_w)$ as defined by Eq. (\ref{viol-coarse}) instead
of our original definition, one can easily prove that \emph{all} the results we have derived in this paper continue
to be valid.

Our analysis is different than the kind of study currently considered in the literature. Instead of focusing
on the statistical properties of the final value of the stochastic entropy production at the end of a given protocol,
we have considered the whole evolution of the stochastic entropy production rate within the time interval.
Even when both approaches are clearly different, they are closely related. Consider, for instance, a system with
many degrees of freedom, or in an asymptotic steady regime. In both cases the probability density functions of both,
the violation fraction and of the stochastic entropy production $S$, are very narrowed around their respective means.
In these scenarios, a large-deviation function exists for both quantities. We then have in those cases
\begin{align}
 \label{correspondence-1}
 \langle S(\tau)\rangle &=\int_0^1\langle S(\tau)|\upsilon\rangle\rho(\upsilon,\tau)d\upsilon\approx
 \big\langle S(\tau)|\langle\upsilon(\tau)\rangle\big\rangle,\\
 \label{correspondence-2}
  \langle\upsilon(\tau)\rangle &=\int_{-\infty}^{\infty}\langle\upsilon(\tau)|S\rangle P(S,\tau)dS\approx
 \big\langle\upsilon(\tau)|\langle S(\tau)\rangle\big\rangle,
\end{align}
where the second relations in Eqs. (\ref{correspondence-1}) and (\ref{correspondence-2}) follow from the saddle-point
evaluation of the corresponding integrals. Thus, there is a one to one correspondence between the mean value of the
entropy production and of the violation fraction, i.e., controlling one of these quantities it is possible to control 
the other.

We discuss now a possible application of our results for free energy recovery in single molecule experiments. 
Note that, for a system initially prepared
in a given equilibrium steady-state, and using the definition of $\Phi$, Eq. (\ref{main-2}), we can write
\begin{equation}
 \label{free-energy-recovery-1}
 \Phi(\upsilon,\tau)=\Psi(\upsilon,\tau)-\beta\Delta F(\tau),
\end{equation}
where
\begin{equation}
 \label{free-energy-recovery-2}
 \Psi(\upsilon,\tau)=-\ln\langle e^{-\beta W(\tau)}|\upsilon\rangle,
\end{equation}
and $\Delta F$ is the change of the free energy during the protocol. $\beta$ corresponds
to the inverse temperature. Then, for example, from Eq. (\ref{IFT}), we can write
\begin{equation}
 \label{free-energy-recovery-3}
 \beta\Delta F(\tau)=-\ln\langle e^{-\Psi(\upsilon,\tau)}\rangle.
\end{equation}
An even more precise method is to consider Eq. (\ref{free-energy-recovery-1}). Using the fact
that $\Phi(1/2,\tau)=0$, we have
\begin{equation}
 \label{free-energy-recovery-4}
 \beta\Delta F(\tau)=\Psi(1/2,\tau).
\end{equation}
The conceptual problem is that one needs to unmask the behavior of the function $\Phi$ (and correspondingly, of $\Psi$),
but the advantage comes from the experimental (or computational) side. 
Note that $\upsilon$ only depends on the instantaneous sign
of the entropy production rate, i.e., one does not need its value and it is sufficient to measure the relative
orientation of a velocity with respect to a probability current (this determines, quite generally, the sign of $\sigma$).
Although one still needs to measure $W$ at the end of the interval in order to have an independent
measure of $\Psi$ (c.f. Eq. (\ref{free-energy-recovery-2})), 
adding the violation fraction in the analysis could help to reduce the error in the estimation
of free energies from single molecule experiments. 
We believe that the discussion above is interesting enough as to motivate the study of the statistics of the violation
fraction in more detail. 

Although the analytical treatement of these problems may prove to be hard, there is a lot of accumulated knowledge
we can borrow from the study of the zero-crossing properties of 
generic stochastic processes. This kind of study
could open the door to new and fruitful collaborations between different fields of statistical mechanics.

In conclusion, we have studied the statistics of the occurrence of entropy-consuming events for single
trajectories of processes in phase space.  We were able to obtain a symmetry relation for the duration
of these events, which is reminiscent of fluctuation theorems and which involves an asymmetry function which
have been studied and characterized within this work. We have studied analytically the steady-state limit
of this symmetry for a paradigmatic model system, showing that even in the simplest cases it is difficult
to say much analytically. However, we believe, as discused above, that our study could be of experimental 
(and computational) relevance, for instance, for the free energy recovery in single molecule experiments.

\begin{acknowledgments}
This work was supported by CNEA, CONICET (PIP11220090100051), and ANPCYT (PICT2011-1537).
R.G.G. thanks V. Lecomte for valuable comments during the early stages of preparation of the present
manuscript, G. Schehr and A. Rosso for providing 
Refs.~\cite{densing2012occupation,majumdar2002large,bray2013persistence}, and Y. N\'u\~{n}ez-Fern\'andez
for fruitful discussions.
\end{acknowledgments}

\bibliographystyle{apsrev4-1.bst}
\bibliography{Violation-3-1}
\end{document}